\documentclass[12pt]{article}

\begin{document}

\title{\bf \Large Symmetries of the Energy-Momentum Tensor of Spherically Symmetric
Lorentzian Manifolds}

\author{M. Sharif \thanks{Present Address: Department of Mathematical Sciences,
University of Aberdeen, Kings College, Aberdeen AB24 3UE Scotland,
UK. $<$msharif@maths.abdn.ac.uk$>$}
\\ Department of Mathematics, University of the Punjab,\\ Quaid-e-Azam Campus
Lahore-54590, PAKISTAN.}

\date{}

\maketitle

\begin{abstract}
Matter collineations of spherically Symmetric Lorentzian Manifolds
are considered. These are investigated when the energy-momentum
tensor is non-degenerate and also when it is degenerate. We have
classified spacetimes admitting higher symmetries and spacetimes
admitting $SO(3)$ as the maximal isometry group. For the
non-degenerate case, we obtain either {\it four}, {\it six}, {\it
seven} or {\it ten} independent matter collineations in which {\it
four} are isometries and the rest are proper. The results of the
previous paper [1] are recovered as a special case. It is worth
noting that we have also obtained two cases where the
energy-momentum tensor is degenerate but the group of matter
collineations is finite-dimensional, i.e. {\it four} or {\it ten}.
\end{abstract}


\date{}

\newpage

\section{Introduction}

Since the pioneering work of Katzin, Levine, Davis and their
collaborators [2]-[6], the study of symmetries played an important
role in the classification of spacetimes, giving rise to many
interesting results with useful applications. The theory of
General Relativity (GR), described by the Einstein's field
equations (EFEs), is a highly non-linear. Due to its
non-linearity, it becomes difficult to find the exact solutions of
the EFEs, in particular, if the metric depends on all coordinates
[7]. However, this problem can be overcome to some extent if it is
assumed that the spacetime has some geometric symmetry properties.
These symmetry properties are given by Killing vectors (KVs) which
then lead to conservation laws [8]-[10]. A large number of
solutions of the EFEs with different symmetry structures have been
found [9] and classified according to their properties [11].

As given by the pioneers, curvature and Ricci tensors play a
significant role (in terms of curvature and Ricci collineations)
in understanding the geometric structure of metrics. They have
provided a detailed study of curvature and Ricci collineations in
the context of the related particle and field conservation laws.
For a given distribution of matter, the contribution of
gravitational potential satisfying EFEs is the principal aim of
all investigations in gravitational physics. This has been
achieved by imposing symmetries on the geometry compatible with
the dynamics of the chosen distribution of matter. In an attempt
to study the geometric and physical properties of the
electromagnetic fields, different types of collineations have been
investigated [12,13] along with many other interesting results.
Symmetries of the energy-momentum tensor (also called matter
collineations) provide conservation laws on matter fields. These
enable us to know how the physical fields, occupying in certain
region of spacetimes, reflect the symmetries of the metric [14].

There is a large body of recent literature which shows interest in
the study of MCs [1],[15]-[22]. In a recent paper [1], the study
of MCs has been taken for static spherically symmetric spacetimes
(SSS) and some interesting results have been obtained. However, it
was incomplete in the sense that (i) only the static case was
considered and (ii) some cases were missing, in particular, for
finite-dimensional MCs. In this paper, we extend the procedure to
calculate MCs of SSS both for non-degenerate and also for
degenerate cases with special emphasis of the metrics admitting
higher symmetries and also $SO(3)$ as the maximal symmetry. We
relate them with RCs and isometries. The rest of the paper is
organized as follows. The next section contains a brief review of
MCs and we write down MC equations for SSS. In section 3, we shall
solve these MC equations when the energy-momentum tensor is
non-degenerate and in the next section MC equations are solved for
the degenerate energy-momentum tensor. Section 5 contains a
summary and discussion of the results obtained.

\section{Matter Collineations and its Equations}

Let $(M,g)$ be a spacetime, where $M$ is a smooth, connected,
Hausdorff four-dimensional manifold and $g$ is smooth Lorentzian
metric of signature (+ - - -) defined on $M$. The manifold $M$ and
the metric $g$ are assumed smooth ($C^{\infty}$). We shall use the
usual component notation in local charts, and a covariant
derivative with respect to the symmetric connection $\Gamma$
associated with the metric $g$ will be denoted by a semicolon and
a partial derivative by a comma.

The geometry and matter of a spacetime are related through the
EFEs given in each coordinate system of $M$ by
\begin{equation}
R_{ab}-\frac{1}{2}Rg_{ab}\equiv G_{ab}=\kappa T_{ab}, \quad
(a,b=0,1,2,3),
\end{equation}
where $\kappa$ is the gravitational constant, $G_{ab}$ is the
Einstein tensor, $R_{ab}$ is the Ricci and $T_{ab}$ is the matter
(energy-momentum) tensor. Also, $R = g^{ab} R_{ab}$ is the Ricci
scalar. We have assumed here that the cosmological constant
$\Lambda=0$. Using the Bianchi identities, it can easily be shown
that
\begin{equation}
G^{ab}_{}{}{;b}=0~~(\Leftrightarrow T^{ab}_{}{}{;b=0}).
\end{equation}
A smooth vector field ${\bf \xi}$ is said to preserve a matter
symmetry [23] on $M$ if, for each smooth local diffeomorphism
$\phi_t$ associated ${\bf \xi}$, the tensor $T$ and $\phi^*_tT$
are equal on the domain $U$ of $\phi_t$, i.e., $T=\phi_t^*T$.
Equivalently, a vector field $\xi^a$ is said to generate a matter
collineation (MC) if it satisfies the following equation
\begin{equation}
\pounds_{\xi}T_{ab}=0~~\Leftrightarrow~~\pounds_{\xi}G_{ab}=0,
\end{equation}
where $\pounds$ is the Lie derivative operator, $\xi^a$ is the
symmetry or collineation vector. Every KV is an MC but the
converse is not true, in general. Collineations can be proper
(non-trivial) or improper (trivial). We define a proper MC to be
an MC which is not a KV, or a homothetic vector (HV). The MC
Eq.(4) can be written in component form as
\begin{equation}
T_{ab,c} \xi^c + T_{ac} \xi^c_{,b} + T_{cb} \xi^c_{,a} = 0.
\end{equation}

The most general form of the metric for a spherically symmetric
Lorentzian manifold is given by
\begin{equation}
ds^2=e^{\nu(t,r)}dt^2-e^{\mu(t,r)}dr^2-e^{\lambda(t,r)}d\Omega^2,
\end{equation}
where $d\Omega^2=d\theta^2+\sin^2\theta d\phi^2$. The surviving
components of the energy-momentum tensor, given in Appendix A, are
$T_{00},~T_{01}~T_{11},~T_{22},~T_{33}$, where
$T_{33}=\sin^2\theta T_{22}$.

The MC equations can be written as follows
\begin{equation}
T_{00,0}\xi^0+T_{00,1}\xi^1+2T_{00}\xi^0_{,0}
+2T_{01}\xi^1_{,0}=0,
\end{equation}
\begin{equation}
T_{01,0}\xi^0 + T_{01,1}\xi^1+T_{01}\xi^0_{,0}+T_{11}\xi^1_{,0}
+T_{01}\xi^1_{,1}+T_{00}\xi^0_{,1}=0,
\end{equation}
\begin{equation}
T_{00}\xi^0_{,2}+T_{01}\xi^1_{,2}+T_{22}\xi^2_{,0}=0,
\end{equation}
\begin{equation}
T_{00}\xi^0_{,3}+T_{01}\xi^1_{,3}+\sin^2\theta T_{22}\xi^3_{,0}=0,
\end{equation}
\begin{equation}
T_{11,0}\xi^0+T_{11,1}\xi^1+2T_{01}\xi^0_{,1}+2T_{11}\xi^1_{,1}=0,
\end{equation}
\begin{equation}
T_{01}\xi^0_{,2}+T_{11}\xi^1_{,2}+T_{22}\xi^2_{,1}=0,
\end{equation}
\begin{equation}
T_{01}\xi^0_{,3}+T_{11}\xi^1_{,3}+\sin^2\theta T_{22}\xi^3_{,1}=0,
\end{equation}
\begin{equation}
T_{22,0}\xi^0+T_{22,1}\xi^1+2T_{22}\xi^2_{,2}=0,
\end{equation}
\begin{equation}
T_{22}\xi^2_{,3}+\sin^2\theta T_{22}\xi^3_{,2}=0,
\end{equation}
\begin{equation}
T_{22,0}\xi^0 + T_{22,1}\xi^1+2\cot\theta
T_{22}\xi^2+2T_{22}\xi^3_{,3} = 0.
\end{equation}
These are the first order non-linear partial differential
equations in four variables $\xi^a(x^b)$. We solve these equations
for the non-degenerate case, when
\begin{equation}
\det(T_{ab})=T^2_{22}(T_{00}T_{11}-T^2_{01})\sin^2\theta\neq 0
\end{equation}
and for the degenerate case, where $\det(T_{ab})=0$. It is noticed
that when $T_{01}=0$ we shall use the notation $T_{aa}=T_a$ for
the sake of brevity.

\section{Matter Collineations in the Non-Degenerate Case}

In this section, we shall evaluate MCs only for those cases which
have non-degenerate energy-momentum tensor, i.e.,
$\det(T_{ab})\neq 0$. This will be done as two cases; one when $M$
admits higher symmetries and one when $SO(3)$ is the maximal
isometry group of $M$. To this end, we set up the general
conditions for the solution of MC equations for the non-degenerate
case.

When we solve Eqs.(6)-(15) simultaneously, after some tedious
algebra, we get the following solution
\begin{eqnarray}
\xi^0=\frac{T_{22}}{T_{00}T_{11}-T^2_{01}}[\{(\dot{A}_1T_{11}
-A'_1T_{01})\sin\phi-(\dot{A}_2T_{11}
-A'_2T_{01})\cos\phi\}\sin\theta\nonumber\\
+(\dot{A}_3T_{11} -A'_3T_{01})\cos\theta+A_4T_{11}-A_5T_{01}],
\end{eqnarray}
\begin{eqnarray}
\xi^1=\frac{-T_{22}}{T_{00}T_{11}-T^2_{01}}[\{(\dot{A}_1T_{01}
-A'_1T_{00})\sin\phi-(\dot{A}_2T_{01}
-A'_2T_{00})\cos\phi\}\sin\theta\nonumber\\
+(\dot{A}_3T_{01} -A'_3T_{00})\sin\theta+A_4T_{01}-A_5T_{00}].
\end{eqnarray}
\begin{eqnarray}
\xi^2=-(A_1\sin\phi-A_2\cos\phi)\cos\theta+A_3\sin\theta\nonumber\\
+c_1\sin\phi -c_2\cos\phi+c_4\ln(\tan\frac{\theta}{2})\sin\theta,
\end{eqnarray}
\begin{eqnarray}
\xi^3=-(A_1\cos\phi+A_2\sin\phi)\csc\theta +(c_1\cos\phi
+c_2\sin\phi)\cot\theta+c_4\phi+c_3.
\end{eqnarray}
where $c_1,c_2,c_3,c_4$ are arbitrary constants and
$A_\nu=A_\nu(t,r),~\nu=1,2,3,4,5$. Here dot and prime indicate the
differentiation with respect to time  and $r$ coordinate
respectively. When we replace these values of $\xi^a$ in MC
Eqs.(6)-(15), we obtain the following differential constraints on
$A_\nu$ with $c_4=0$
\begin{eqnarray}
2(T_{00}T_{11}-T^2_{01})(T_{22}\dot{A}_i\dot{)}
+T_{22}[(2T_{01}\dot{T}_{01}-T_{11}\dot{T}_{00}
-T_{01}T'_{00})\dot{A}_i\nonumber\\
-(2T_{00}\dot{T}_{01}
-T_{01}\dot{T}_{00}-T_{00}T'_{00})A'_i]=0,~~(i=1,2,3),
\end{eqnarray}
\begin{eqnarray}
(T_{00}T_{11}-T^2_{01})[(T_{22}\dot{A}_i)'+(T_{22}A'_i\dot{)}]
+T_{22}[(T_{01}\dot{T}_{11}-T_{11}T'_{00})\dot{A}_i\nonumber\\
+(T_{01}T'_{00} -T_{00}\dot{T}_{11})A'_i]=0,
\end{eqnarray}
\begin{eqnarray}
2(T_{00}T_{11}-T^2_{01})(T_{22}A'_i)'
+T_{22}[(2T_{01}T'_{01}-T_{00}T'_{11}
-T_{01}\dot{T}_{11})\nonumber\\
-(2T_{11}T'_{01} -T_{11}\dot{T}_{11}-T_{01}T'_{11})\dot{A}_i]=0,
\end{eqnarray}
\begin{eqnarray}
(T_{11}\dot{T}_{22}-T_{01}T'_{22})\dot{A}_i+(T_{00}{T'}_{22}
-T_{01}\dot{T}_{22})A'_i+2A_i=0,
\end{eqnarray}
\begin{eqnarray}
2(T_{00}T_{11}-T^2_{01})(T_{22}A_4\dot{)}
+T_{22}[(2T_{01}\dot{T}_{01}-T_{11}\dot{T}_{00}
-T_{01}T'_{00})A_4\nonumber\\
-(2T_{00}\dot{T}_{01} -T_{01}\dot{T}_{00}-T_{00}T'_{00})A_5]=0,
\end{eqnarray}
\begin{eqnarray}
(T_{00}T_{11}-T^2_{01})[(T_{22}A_4)'+(T_{22}A_5\dot{)}]
+T_{22}[(T_{01}\dot{T}_{11}-T_{11}T'_{00})A_4\nonumber\\
+(T_{01}T'_{00} -T_{00}\dot{T}_{11})A_5]=0,
\end{eqnarray}
\begin{eqnarray}
2(T_{00}T_{11}-T^2_{01})(T_{22}A_5)'
+T_{22}[(2T_{01}T'_{01}-T_{00}T'_{11}
-T_{01}\dot{T}_{11})A_5\nonumber\\
-(2T_{11}T'_{01}-T_{11}\dot{T}_{11}-T_{01}T'_{11})A_4]=0,
\end{eqnarray}
\begin{eqnarray}
(T_{11}\dot{T}_{22}-T_{01}T'_{22})A_4+(T_{00}{T'}_{22}
-T_{01}\dot{T}_{22})A_5=0.
\end{eqnarray}
Thus the problem of working out MCs for all possibilities of
$A_i,A_4,A_5$ is reduced to solving the set of Eqs.(17)-(20)
subject to the above constraints. We would solve these to classify
MCs of the manifolds admitting higher symmetries than $SO(3)$ and
$SO(3)$ as the maximal isometry group.

\subsection{MCs of the Spacetimes Admitting Higher Symmetries}

Here we use the constraint Eqs.(21)-(28) to evaluate MCs of the
spacetimes given by Eq.(5) which admit higher symmetries than
$SO(3)$. The six cases admitting symmetry groups larger
than $SO(3)$ are the following:\\
\par \noindent
\par \noindent
(1) $~~~~~SO(3)\otimes{\bf R}$, where ${\bf R}=\partial_t$ if and
only if
\par \noindent
\par \noindent
(a) $~~~~~\nu=\nu(r),~\mu=\mu(r),~\lambda=2\ln r$ or (b)
$~~~~~\nu=\nu(r),~\mu=0,~\lambda=2\ln a$,
\par \noindent
\par \noindent
$~~~~~~~~~~$where $a$ is an arbitrary constant,\\
\par \noindent
\par \noindent
(2) $~~~~~SO(3)\otimes{\bf R}$, where ${\bf R}=\partial_r$ if and
only if
\par \noindent
\par \noindent
(a) $~~~~~\nu=\nu(t),~\mu=\mu(t),~\lambda=2\ln t$ or (b)
$~~~~~\nu=0,~\mu=\mu(t),~\lambda=2\ln a$,\\
\par \noindent
\par \noindent
(3) $~~~~~SO(3)\otimes{\bf R}$, where ${\bf R}
=\partial_t+e\partial_r$ if and only if
\par \noindent
\par \noindent
$~~~~~~~~~~\nu=0=\mu,~\lambda=\lambda(t+er)$ with $e=\pm 1$,
\par \noindent
\par \noindent\\
(4) $~~~~~SO(4)$ if and only if $\nu=0,~\mu=2\ln
R(t),~\lambda=2\ln R(t)\sin r$ such that
\par \noindent
\par \noindent
$~~~~~~~~~~R\ddot{R}-\dot{R}^2-1\neq 0$,\\
\par \noindent
\par \noindent
(5) $~~~~~SO(3)\times {\bf R^3}$ if and only if $\nu=0,~\mu=2\ln
R(t),~\lambda=2\ln R(t)r$ such $~~~~~~~~~~$ that
\par \noindent
\par \noindent
$~~~~~~~~~~R\ddot{R}-\dot{R}^2\neq 0$,\\
\par \noindent
\par \noindent
(6) $~~~~~SO(1,3)$ if and only if
\par \noindent
\par \noindent
(a) $~~~~~\nu=0,~\mu=2\ln R(t),~\lambda=2\ln R(t)\sinh r$ such
that
\par \noindent
\par \noindent
$~~~~~~~~~~R\ddot{R}-\dot{R}^2+1\neq 0$, or
\par \noindent
\par \noindent
(b) $~~~~~\nu=2\ln Q(r),~\mu=0,~\lambda=2\ln Q(r)\cosh t$ such
that
\par \noindent
\par \noindent
$~~~~~~~~~~QQ''-Q'^2+1\neq 0$.\\
\par \noindent
\par \noindent
{\bf Case (1)}: In this case, we have $T_{01}=0$ and also
$\dot{T}_{ab}=0$. Using these values, Eqs.(17)-(28) reduce to
\begin{eqnarray}
\xi^0&=&\frac{T_2}{T_0}[(\dot{A}_1\sin\phi
-\dot{A}_2\cos\phi)\sin\theta+\dot{A}_3\cos\theta+A_4],\\
\xi^1&=&\frac{T_2}{T_1}[(A_1'\sin\phi
-A'_2\cos\phi)\sin\theta+A'_3\cos\theta+A_5],\\
\xi^2&=&-(A_1\sin\phi-A_2\cos \phi)\cos\theta+A_3\sin\theta+
c_1\sin\phi-c_2\cos\phi,\\
\xi^3&=&-(A_1\cos\phi+A_2\sin\phi)\csc\theta+(c_1\cos\phi
+c_2\sin\phi)\cot\theta+c_3 ,
\end{eqnarray}
where we have used the notation $T_{aa}=T_a$ for the sake of
simplicity. These $\xi^a$ are satisfied subject to the following
differential constraints on $A_\nu$
\begin{eqnarray}
T_1\dot{A}_4+T'_0A_5=0,~~(\frac{T_2}{T_0}A_4)'
+\frac{T_2}{T_0}\dot{A}_5=0,\nonumber\\
(\frac{T_2}{\sqrt{T_1}}A_5)'=0,~~T'_2A_5=0,\\
2T_1\ddot{A}_i+T'_0A'_i=0,~~(\sqrt{\frac{T_2}{T_0}}\dot{A}_i)'=0,\nonumber\\
(\frac{T_2}{\sqrt{T_1}}A'_i)'=0,~~2T_1A_i+T'_2A'_i=0.
\end{eqnarray}
It is interesting to note that this case reduces to the
non-degenerate case of the paper [1]. However, the possibility of
seven MCs is recovered here which was missing there. Now the
evaluation of MCs for all possibilities of $A_i,A_4,A_5$ is
reduced to solving the set of Eqs.(29)-(32) subject to the
constraints given by Eqs.(33) and (34). A complete solution of
these equations is obtained by considering different possibilities
of $T_2$. The last equation of Eq.(33) implies that either
\begin{eqnarray*}
(a)~T'_2=0,~~or~~(b)~T'_2\neq 0.
\end{eqnarray*}

The first case when $T_2=\beta$, where $\beta$ is an arbitrary
constant, Eq.(34) gives $A_i=0$ and consequently Eqs.(29)-(32)
yield
\begin{eqnarray}
\xi^0=A_4(t,r),~\xi^1=A_5(t,r),\nonumber\\
\xi^2=c_1\sin\phi-c_2\cos\phi,~
\xi^3=(c_1\cos\phi+c_2\sin\phi)\cot\theta+c_3.
\end{eqnarray}
Further, if we assume that
$[\frac{T_0}{\sqrt{T_1}}(\frac{T'_0}{2T_0\sqrt{T_1}})']'\neq 0$,
we obtain four MCs identical to the usual KVs of spherical
symmetry given by
\begin{equation}
\xi=c_0\frac{T_0}{\beta}\partial_t+c_1(\sin\phi\partial_\theta
+\cot\theta\cos\phi\partial_\phi)
+c_2(\cos\phi\partial_\theta-\cot\theta\sin\phi\partial_\phi)
+c_3\partial_\phi.
\end{equation}
When $[\frac{T_0}{\sqrt{T_1}}(\frac{T'_0}{2T_0\sqrt{T_1}})']'=0$,
this implies that
$\frac{T_0}{\sqrt{T_1}}[\frac{T'_0}{2T_0\sqrt{T_1}}]'=\alpha$,
where $\alpha$ is an arbitrary constant which may be positive,
zero or negative. In each case, we have six MCs.
\par \noindent
\par \noindent
For $\alpha>0$, we obtain
\begin{eqnarray}
\xi&=&c_0\frac{T_0}{\beta}\partial_t+c_1(\sin\phi\partial_\theta
+\cot\theta\cos\phi\partial_\phi)
+c_2(\cos\phi\partial_\theta-\cot\theta\sin\phi\partial_\phi)
+c_3\partial_\phi\nonumber\\
&+&c_4(-\frac{T'_0}{2\beta\sqrt{\alpha
T_1}}\sinh\sqrt{\alpha}t\partial_t+\frac{\sqrt{T_1}}{\beta}
\cosh\sqrt{\alpha}t\partial_r)\nonumber\\
&+&c_5(-\frac{T'_0}{2\beta\sqrt{\alpha
T_1}}\cosh\sqrt{\alpha}t\partial_t+\frac{\sqrt{T_1}}
{\beta}\sinh\sqrt{\alpha}t\partial_r).
\end{eqnarray}
If $\alpha=0$, we have
\begin{eqnarray}
\xi&=&c_0\frac{T_0}{\beta}\partial_t+c_1(\sin\phi\partial_\theta
+\cot\theta\cos\phi\partial_\phi)
+c_2(\cos\phi\partial_\theta-\cot\theta\sin\phi\partial_\phi)
+c_3\partial_\phi\nonumber\\
&+&c_4[-\frac{T_0}{\beta}(\frac{\gamma}{2}t^2
+\frac{1}{\beta}\int{\sqrt{T_1}}dr)\partial_t
+\frac{\sqrt{T_1}}{\beta}t\partial_r]\nonumber\\
&+&c_5(\frac{T_0}{\beta}\gamma t\partial_t
+\frac{\sqrt{T_1}}{\beta}\partial_r),
\end{eqnarray}
where $\frac{T'_0}{2T_0\sqrt{T_1}}=\gamma$, an arbitrary constant.
The case $\alpha<0$ yields
\begin{eqnarray}
\xi&=&c_0\frac{T_0}{\beta}\partial_t+c_1(\sin\phi\partial_\theta
+\cot\theta\cos\phi\partial_\phi)
+c_2(\cos\phi\partial_\theta-\cot\theta\sin\phi\partial_\phi)
+c_3\partial_\phi\nonumber\\
&+&c_4(-\frac{T'_0}{2\beta\sqrt{-\alpha
T_1}}\sin\sqrt{-\alpha}t\partial_t+\frac{\sqrt{T_1}}{\beta}
\cos\sqrt{-\alpha}t\partial_r)\nonumber\\
&+&c_5(\frac{T'_0}{2\beta\sqrt{-\alpha
T_1}}\cos\sqrt{-\alpha}t\partial_t+\frac{\sqrt{T_1}}
{\beta}\sin\sqrt{-\alpha}t\partial_r).
\end{eqnarray}

In the case (b), when $T'_2\neq 0$, it follows from Eqs.(33) and
(34) that for
$\frac{T_2}{\sqrt{T_1}}(\frac{T'_2}{2T_2\sqrt{T_1}})'+1\neq 0$, we
obtain the same MCs as the usual minimal KVs for spherically
symmetry.

If $\frac{T_2}{\sqrt{T_1}}(\frac{T'_2}{2T_2\sqrt{T_1}})'+1=0$ and
$(\frac{T'_2}{\sqrt{T_0T_1T_2}})'\neq 0$, we have seven MCs given
by
\begin{eqnarray}
\xi&=&c_0\partial_t+c_1(\sin\phi\partial_\theta
+\cot\theta\cos\phi\partial_\phi)
+c_2(\cos\phi\partial_\theta-\cot\theta\sin\phi\partial_\phi)
+c_3\partial_\phi\nonumber\\
&+&c_4(-\frac{1}{\sqrt{T_1}}\sin\phi\sin\theta\partial_r
-X\sin\phi\cos\theta\partial_\theta
-X\cos\phi\csc\theta\partial_\phi)\nonumber\\
&+&c_5(\frac{1}{\sqrt{T_1}}\cos\phi\sin\theta\partial_r
+X\cos\phi\cos\theta\partial_\theta
-X\sin\phi\csc\theta\partial_\phi)\nonumber\\
&+&c_6(-\frac{1}{\sqrt{T_1}}\cos\theta\partial_r
-X\sin\theta\partial_\theta),
\end{eqnarray}
where $X=\frac{T'_2}{2T_2\sqrt{T_1}}$. If we have
$\frac{T_2}{\sqrt{T_1}}(\frac{T'_2}{2T_2\sqrt{T_1}})'+1=0,~
(\frac{T'_2}{\sqrt{T_0T_1T_2}})'=0$ and $(\frac{T'_0}{T'_2})'\neq
0$, then we get four MCs.

When $\frac{T_2}{\sqrt{T_1}}(\frac{T'_2}{2T_2\sqrt{T_1}})'+1=0,~
(\frac{T'_2}{\sqrt{T_0T_1T_2}})'=0$ and
$\frac{T'_0}{T'_2}=\delta$, an arbitrary constant. For $\delta>0$,
we obtain
\begin{eqnarray}
\xi&=&c_0\partial_t+c_1(\sin\phi\partial_\theta
+\cot\theta\cos\phi\partial_\phi)
+c_2(\cos\phi\partial_\theta-\cot\theta\sin\phi\partial_\phi)
+c_3\partial_\phi\nonumber\\
&+&c_4[(\frac{T_2}{T_0}X\sqrt{\delta}\sinh\sqrt{\delta}t\partial_t
-\frac{1}{\sqrt{T_1}}\cosh\sqrt{\delta}t\partial_r)\sin\theta\sin\phi\nonumber\\
&-&(\cos\theta\sin\phi\partial_\theta+\csc\theta\cos\phi\partial_\phi)
X\cosh\sqrt{\delta}t]\nonumber\\
&+&c_5[(-\frac{T_2}{T_0}X\sqrt{\delta}\sinh\sqrt{\delta}t\partial_t
+\frac{1}{\sqrt{T_1}}\cosh\sqrt{\delta}t\partial_r)\sin\theta\cos\phi\nonumber\\
&+&(\cos\theta\cos\phi\partial_\theta-\csc\theta\sin\phi\partial_\phi)
X\cosh\sqrt{\delta}t]\nonumber\\
&+&c_6[(\frac{T_2}{T_0}X\sqrt{\delta}\sinh\sqrt{\delta}t\partial_t
-\frac{1}{\sqrt{T_1}}\cosh\sqrt{\delta}t\partial_r)\cos\theta
+X\cosh\sqrt{\delta}t\sin\theta\partial_\theta]\nonumber\\
&+&c_7[(\frac{T_2}{T_0}X\sqrt{\delta}\cosh\sqrt{\delta}t\partial_t
-\frac{1}{\sqrt{T_1}}\sinh\sqrt{\delta}t\partial_r)\sin\theta\sin\phi\nonumber\\
&-&(\cos\theta\sin\phi\partial_\theta+\csc\theta\cos\phi\partial_\phi)
X\sinh\sqrt{\delta}t]\nonumber\\
&+&c_8[-(\frac{T_2}{T_0}X\sqrt{\delta}\cosh\sqrt{\delta}t\partial_t
+\frac{1}{\sqrt{T_1}}\sinh\sqrt{\delta}t\partial_r)\sin\theta\cos\phi\nonumber\\
&+&(\cos\theta\cos\phi\partial_\theta-\csc\theta\sin\phi\partial_\phi)
X\sinh\sqrt{\delta}t\nonumber\\
&+&c_9(\frac{T_2}{T_0}X\sqrt{\delta}\cosh\sqrt{\delta}t\partial_t
-\frac{1}{\sqrt{T_1}}X\sinh\sqrt{\delta}t\partial_r)\cos\theta
+X\sinh\sqrt{\delta}t\sin\theta\partial_\theta].\nonumber\\
\end{eqnarray}
If $\delta=0$, we have
\begin{eqnarray}
\xi&=&c_0\partial_t+c_1(\sin\phi\partial_\theta
+\cot\theta\cos\phi\partial_\phi)
+c_2(\cos\phi\partial_\theta-\cot\theta\sin\phi\partial_\phi)
+c_3\partial_\phi\nonumber\\
&+&c_4[(\frac{T_2}{T_0}X\partial_t
-\frac{1}{\sqrt{T_1}}t\partial_r)\sin\theta\sin\phi
-(\cos\theta\sin\phi\partial_\theta+\csc\theta\cos\phi\partial_\phi)
tX]\nonumber\\
&+&c_5[(-\frac{T_2}{T_0}X\partial_t
+\frac{1}{\sqrt{T_1}}t\partial_r)\sin\theta\cos\phi
+(\cos\theta\cos\phi\partial_\theta-\csc\theta\sin\phi\partial_\phi)
tX]\nonumber\\
&+&c_6[(\frac{T_2}{T_0}X\partial_t
-\frac{1}{\sqrt{T_1}}t\partial_r)\cos\theta+tX\sin\theta]\nonumber\\
&+&c_7[(-\frac{1}{\sqrt{T_1}}\sin\theta\partial_r
-X\cos\theta\partial_\theta)\sin\phi-X\csc\theta\cos\phi\partial_\phi]\nonumber\\
&+&c_8[(\frac{1}{\sqrt{T_1}}\sin\theta\partial_r
+X\cos\theta\partial_\theta)\cos\phi-X\csc\theta\sin\phi\partial_\phi]\nonumber\\
&+&c_9(-\frac{1}{\sqrt{T_1}}\cos\theta\partial_r+X\sin\theta\partial_\theta).
\end{eqnarray}
For $\delta<0$, MCs are given by
\begin{eqnarray}
\xi&=&c_0\partial_t+c_1(\sin\phi\partial_\theta
+\cot\theta\cos\phi\partial_\phi)
+c_2(\cos\phi\partial_\theta-\cot\theta\sin\phi\partial_\phi)
+c_3\partial_\phi\nonumber\\
&+&c_4[(-\frac{T_2}{T_0}X\sqrt{-\delta}\sin\sqrt{-\delta}t\partial_t
-\frac{1}{\sqrt{T_1}}\cos\sqrt{-\delta}t\partial_r)\sin\theta\sin\phi\nonumber\\
&-&(\cos\theta\sin\phi\partial_\theta+\csc\theta\cos\phi\partial_\phi)
X\cos\sqrt{-\delta}t]\nonumber\\
&+&c_5[(\frac{T_2}{T_0}X\sqrt{-\delta}\sin\sqrt{-\delta}t\partial_t
+\frac{1}{\sqrt{T_1}}\cos\sqrt{-\delta}t\partial_r)\sin\theta\cos\phi\nonumber\\
&+&(\cos\theta\cos\phi\partial_\theta-\csc\theta\sin\phi\partial_\phi)
X\cos\sqrt{-\delta}t]\nonumber\\
&+&c_6[(-\frac{T_2}{T_0}X\sqrt{-\delta}\sin\sqrt{-\delta}t\partial_t
-\frac{1}{\sqrt{T_1}}\cos\sqrt{-\delta}t\partial_r)\cos\theta
+X\cos\sqrt{-\delta}t\sin\theta\partial_\theta]\nonumber\\
&+&c_7[(\frac{T_2}{T_0}X\sqrt{-\delta}\cos\sqrt{-\delta}t\partial_t
-\frac{1}{\sqrt{T_1}}\sin\sqrt{-\delta}t\partial_r)\sin\theta\sin\phi\nonumber\\
&-&(\cos\theta\sin\phi\partial_\theta+\csc\theta\cos\phi\partial_\phi)
X\sin\sqrt{-\delta}t]\nonumber\\
&+&c_8[-(\frac{T_2}{T_0}X\sqrt{\delta}\cos\sqrt{-\delta}t\partial_t
+\frac{1}{\sqrt{T_1}}\sin\sqrt{-\delta}t\partial_r)\sin\theta\cos\phi\nonumber\\
&+&(\cos\theta\cos\phi\partial_\theta-\csc\theta\sin\phi\partial_\phi)
X\sin\sqrt{-\delta}t\nonumber\\
&+&c_9(\frac{T_2}{T_0}X\sqrt{-\delta}\cos\sqrt{-\delta}t\partial_t
-\frac{1}{\sqrt{T_1}}X\sin\sqrt{-\delta}t\partial_r)\cos\theta
+X\sin\sqrt{-\delta}t\sin\theta\partial_\theta].\nonumber\\
\end{eqnarray}
From Eqs.(41)-(43), it follows that for each value of $\delta$, we
obtain ten independent MCs.
\par \noindent
\par \noindent
{\bf Case (2)}: In this case, we have $T_{01}=0$ and $T'_a=0$. If
we use the transformations $t\leftrightarrow
r,~\xi^0\leftrightarrow \xi^1,~T_0\leftrightarrow T_1$, the
solution of this case can be trivially obtained as in the case
(1).
\par \noindent
\par \noindent
{\bf Cases (4), (5), (6)}: The cases (4), (5) and (6a) describe
Friedmann Robertson (FRW) spacetimes where as the case (6b)
describes FRW like spacetimes. For these metrics, the
non-vanishing components of Ricci and energy-momentum tensors are
given in Appendix B. If any of $T_a$ is zero, we get infinte
dimensional MCs. For the non-degenerate case, we have $T_a\neq 0$
which implies the following possibilities
\begin{eqnarray*}
(a)~~~\frac{T_1}{\sqrt{T_0}}(\frac{\dot{T_2}}{2T_1\sqrt{T_0}}\dot{)}
-k=0,~~(b)~~~
\frac{T_1}{\sqrt{T_0}}(\frac{\dot{T_2}}{2T_1\sqrt{T_0}}\dot{)}
-k\neq 0
\end{eqnarray*}
with
\begin{eqnarray*}
(i)~~~\dot{T_1}=0,~~(ii)~~~\dot{T_1}\neq 0,
\end{eqnarray*}
where $k$ has the values $1,0,-1$ according as for closed, flat
and open FRW spacetimes respectively.

In the case (ai), we must have $k=0$ and $T_1=a\neq 0,~a$ is an
arbitrary constant. Thus, in addition to the non-proper MCs
$\xi_{(1)},\xi_{(2)},\xi_{(3)},\xi_{(4)},\xi_{(5)},\xi_{(6)}$
given in Appendix C, we obtain the following proper MCs
\begin{eqnarray}
\xi_{(7)}&=&\frac{1}{\sqrt{T_0}}\partial_t,\nonumber\\
\xi_{(8)}&=&r(\frac{1}{\sqrt{T_0}}\partial_t
-Y\partial_r)\sin\theta\sin\phi-(\cos\theta\sin\phi
\partial_\theta+\csc\theta\cos\phi\partial_\phi)Y,\nonumber\\
\xi_{(9)}&=&r(\frac{1}{\sqrt{T_0}}\partial_t
-Y\partial_r)\sin\theta\cos\phi
-(\cos\theta\cos\phi\partial_\theta
-\csc\theta\sin\phi\partial_\phi)Y,\nonumber\\
\xi_{(10)}&=&r(\frac{1}{\sqrt{T_0}}\partial_t
-Y\partial_r)\cos\theta+Y\sin\theta\partial_\theta.
\end{eqnarray}
where $Y=\frac{1}{ar}\int{\sqrt{T_0}dt}$. This gives ten
independent MCs in which six are the usual KVs of of closed FRW
metric and the rest are the proper MCs.

The case (aii) also yields ten independent MCs for each value of
$k$. For the value of $k=1$, the proper MCs are given by
\begin{eqnarray}
\xi_{(7)}&=&(\frac{\sqrt{T_0}}{T_1}\cot r\partial_t
-Z\sin^2r\partial_r)\csc r,\nonumber\\
\xi_{(8)}&=&[(\frac{T_2}{T_0}\dot{Z}\partial_t
-Z\sin r\cos r\partial_r)\sin\theta\sin\phi\nonumber\\
&-&Z(\cos\theta\sin\phi\partial_\theta
+\csc\theta\cos\phi\partial_\phi)\csc r],\nonumber\\
\xi_{(9)}&=&[(\frac{T_2}{T_0}\dot{Z}\partial_t
-Z\sin r\cos r\partial_r)\sin\theta\cos\phi\nonumber\\
&-&Z(\cos\theta\cos\phi\partial_\theta
-\csc\theta\sin\phi\partial_\phi)\csc r],\nonumber\\
\xi_{(10)}&=&[(\frac{T_2}{T_0}\dot{Z}\partial_t-Z\sin r\cos
r\partial_r)\cos\theta-Z\partial_\theta]\csc r.
\end{eqnarray}
where $Z=\frac{T_2}{2T_1\sqrt{T_0}}$. For $k=0$, we have the
following proper MCs
\begin{eqnarray}
\xi_{(7)}&=&(\frac{1}{\sqrt{T_0}}\partial_t
-rZ\partial_r),\nonumber\\
\xi_{(8)}&=&[\{(\frac{rT_2}{2T_0}\dot{Z}
-\frac{r}{\sqrt{T_0}})\partial_t
+(\frac{r^2Z}{2}+\int{\frac{\sqrt{T_0}}{T_1}}dt
\partial_r)\}\sin\theta\sin\phi,\nonumber\\
&-&(\frac{r}{2}Z-\frac{1}{r}\int{\frac{\sqrt{T_0}}{T_1}}dt)
(\cos\theta\sin\phi\partial_\theta+\csc\theta\cos\phi
\partial_\phi)],\nonumber\\
\xi_{(9)}&=&[\{(\frac{rT_2}{2T_0}\dot{Z}
-\frac{r}{\sqrt{T_0}})\partial_t
+(\frac{r^2Z}{2}+\int{\frac{\sqrt{T_0}}{T_1}}dt
\partial_r)\}\sin\theta\cos\phi\nonumber\\
&-&(\frac{r}{2}Z-\frac{1}{r}\int{\frac{\sqrt{T_0}}{T_1}}dt)
(\cos\theta\cos\phi\partial_\theta-\csc\theta\sin\phi
\partial_\phi)],\nonumber\\
\xi_{(10)}&=&[\{(\frac{rT_2}{2T_0}\dot{Z}
-\frac{r}{\sqrt{T_0}})\partial_t
+(\frac{r^2Z}{2}+\int{\frac{\sqrt{T_0}}{T_1}}dt
\partial_r)\}\cos\theta\nonumber\\
&-&(\frac{r}{2}Z-\frac{1}{r}\int{\frac{\sqrt{T_0}}{T_1}}dt)
\sin\theta\partial_\theta].
\end{eqnarray}
For the value of $k=-1$, the four proper MCs are
\begin{eqnarray}
\xi_{(7)}&=&\frac{1}{T_1}(\sqrt{T_0}\coth r\partial_t
-T_2Z\partial_r)\csc hr,\nonumber\\
\xi_{(8)}&=&[(\frac{T_2}{T_0}\dot{Z}\partial_t
+Z\sinh r\cosh r\partial_r)\sin\theta\sin\phi\nonumber\\
&-&Z(\cos\theta\sin\phi\partial_\theta
+\csc\theta\cos\phi\partial_\phi)]\csc hr,\nonumber\\
\xi_{(9)}&=&[(\frac{T_2}{T_0}\dot{Z}\partial_t
+Z\sinh r\cosh r\partial_r)\sin\theta\cos\phi\nonumber\\
&-&Z(\cos\theta\cos\phi\partial_\theta
-\csc\theta\sin\phi\partial_\phi)]\csc hr,\nonumber\\
\xi_{(10)}&=&[(\frac{T_2}{T_0}\dot{Z}\partial_t +Z\sinh r\cosh
r\partial_r)\cos\theta -Z\partial_\theta]\csc hr.
\end{eqnarray}
Thus we obtain ten independent MCs for each value of $k$ in which
six are the usual isometries of FRW metric and the the remaining
four are the proper MCs.

For the case (bi), we must require that $k\neq 0$. When $k=1$, we
obtain one proper MC given by
\begin{eqnarray}
\xi_{(7)}&=&\frac{T_2}{\sqrt{T_0}a}\csc^2 r\partial_t.
\end{eqnarray}
For $k=-1$, proper MC is
\begin{eqnarray}
\xi_{(7)}&=&\frac{T_2}{\sqrt{T_0}a}\csc h^2 r\partial_t.
\end{eqnarray}
This case gives seven independent MCs in which six are non-proper
and one is proper MC. It can be checked that the case (bii) gives
six independent MCs for each value of $k$ which are usual KVs of
FRW spacetimes. Similarly, the case (6b) can be solved to give
either six, seven or ten MCs.

\subsection{MCs of the Spacetimes Admitting $SO(3)$ as the
Maximal Isometry Group}

In this section, we evaluate MCs of the spherically symmetric
spacetimes which admit $SO(3)$ as the maximal isometry group. In
these solutions, we take any additional MC (if exists) be
orthogonal to the $SO(3)$ orbit. For this we must require that
$A_i\equiv 0$ and consequently, it follows from Eqs.(17)-(20) that
$\xi^0=\xi^0(t,r),~\xi^1=\xi^1(t,r),~\xi^2=0,~ \xi^3=0.$ It is
mentioned here that we are considering only diagonal metrics for
this case. The non-diagonal metrics can be solved in a similar
way. If we make use of the following substitutions
\begin{eqnarray*}
\frac{T_2}{T_0}A_4=C(t,r),~\frac{T_2}{T_1}A_5=D(t,r),
~\sqrt{T_0}=A(t,r),~\sqrt{T_1}=B(t,r)
\end{eqnarray*}
in the constraint Eqs.(21)-(28), then it follows that
\begin{equation}
\dot{C}=-\dot{A}C-A'D,
\end{equation}
\begin{equation}
A^2C'+B^2\dot{D}=0,
\end{equation}
\begin{equation}
D'=-\dot{B}C-B'D,
\end{equation}
\begin{equation}
\dot{T_2}C+T'_2D=0.
\end{equation}
To solve this system of equations, we have the following
possibilities:
\par \noindent
\par \noindent
(i) $~~\dot{T}_2=0,~~T'_2\neq 0,~~~~~$ (ii) $\dot{T}_2\neq
0,~~T'_2=0$,
\par \noindent
\par \noindent
(iii) $\dot{T}_2\neq 0,~~T'_2\neq 0,~~~~~$ (iv)
$\dot{T}_2=0,~~T'_2=0$.

The first possibility does not provide any proper MC if we assume
that $\dot{T}_1\neq 0$. However, the assumption
$\dot{T}_1=0,~\dot{T}_0\neq 0$ gives infinite dimensional MCs.

The second case shows that there does not exist a proper MC with
the constraint $T'_0\neq 0$ but the constraints $T'_0=0,~T'_1\neq
0$ provide infinite dimensional MCs.

In the third case, when
\begin{eqnarray*}
T_0T'_2[\frac{T'_1\dot{T_2}-\dot{T_1}T'_2}{2\sqrt{T_1}\dot{T_2}}
+\{\ln(\frac{\dot{T_2}}{T'_2})\}']
+T_1\dot{T_2}[\frac{T'_0\dot{T_2}-\dot{T_0}T'_2}{2\sqrt{T_0}T'_2}
+\{\ln(\frac{\dot{T_2}}{T'_2})\dot{\}}]\neq 0,
\end{eqnarray*}
we do not have a proper MC. However, if
\begin{eqnarray*}
T_0T'_2[\frac{T'_1\dot{T_2}-\dot{T_1}T'_2}{2\sqrt{T_1}\dot{T_2}}
+\{\ln(\frac{\dot{T_2}}{T'_2})\}']
+T_1\dot{T_2}[\frac{T'_0\dot{T_2}-\dot{T_0}T'_2}{2\sqrt{T_0}T'_2}
+\{\ln(\frac{\dot{T_2}}{T'_2})\dot{\}}]=0.
\end{eqnarray*}
and
\begin{eqnarray*}
(\frac{T'_0\dot{T_2}-\dot{T_0}T'_2}{2\sqrt{T_0}\dot{T_2}})'
=[\frac{T'_2}{\dot{T_2}}\{\frac{\dot{T_1}T'_2-T'_1\dot{T_2}}{2\sqrt{T_1}T'_2}
-(\frac{\dot{T_2}}{T'_2})'\}\dot{]}
\end{eqnarray*}
then there exists a proper MC given by
\begin{eqnarray}
\exp(\int{\frac{T'_0\dot{T_2}-\dot{T_0}T'_2}{2\sqrt{T_0}\dot{T_2}}}dt)
(\partial_t-\frac{\dot{T_2}}{T'_2}\partial_r).
\end{eqnarray}
If
\begin{eqnarray*}
(\frac{T'_0\dot{T_2}-\dot{T_0}T'_2}{2\sqrt{T_0}\dot{T_2}})'
\neq[\frac{T'_2}{\dot{T_2}}\{\frac{\dot{T_1}T'_2-T'_1\dot{T_2}}{2\sqrt{T_1}T'_2}
-(\frac{\dot{T_2}}{T'_2})'\}\dot{]}
\end{eqnarray*}
then this case gives infinite number of MCs.

In the last case, we solve Eqs.(50)-(52) which imply that
$\dot{A}B'-A'\dot{B}\equiv \psi(t,r)$. If $\psi=0$, then we must
have $\dot{C}=0=D'$ for a non-trivial solution. Thus the
constraints $\dot{T}_0T'_1-T'_0\dot{T}_1=0$ together with
\begin{eqnarray*}
T'_0\neq
0,~[\frac{T_1}{T_0}(\frac{\dot{T}_1}{T'_0}\dot{)}\dot{]}=0,
\end{eqnarray*}
yield the following proper MC
\begin{eqnarray}
\exp(\int{\frac{T_1}{T_0}(\frac{\dot{T}_0}{T'_0}\dot{)}}dr)
(\partial_t-\frac{\dot{T}_0}{T'_0}\partial_r).
\end{eqnarray}
However, for $\dot{T}_0T'_1-T'_0\dot{T}_1=0$ together with
\begin{eqnarray*}
T'_0=0,~T'_1\neq
0,~[\frac{T_1}{T_0}(\frac{\dot{T}_1}{T'_1}\dot{)}\dot{]}=0,
\end{eqnarray*}
we obtain the proper MC given by
\begin{eqnarray}
\exp(\int{\frac{T_1}{T_0}(\frac{\dot{T}_1}{T'_1}\dot{)}}dr)
(\partial_t-\frac{\dot{T}_1}{T'_1}\partial_r).
\end{eqnarray}
The constraint $\dot{T}_0T'_1-T'_0\dot{T}_1=0$ along with
$T'_0=0,~T'_1=0=\dot{T}_1,~\dot{T}_0\neq 0$ gives infinite many
MCs.

For $\psi\neq 0$, we must have $\dot{C}\neq 0,D'\neq 0$ for a
non-trivial solution. Let us express $\dot{C}$ and $D'$ as $E$ and
$F$ respectively so that
\begin{equation}
C=-\frac{B'}{\psi}E(t,r)+\frac{A'}{\psi}F(t,r),
\end{equation}
\begin{equation}
D=\frac{\dot{B}}{\psi}E(t,r)-\frac{\dot{A}}{\psi}F(t,r).
\end{equation}
We obtain two linearly independent MCs which are orthogonal to
$T_e(SO(3))$ and are given by
\begin{equation}
X_1=\frac{E}{\psi}(-B'\partial_t+\dot{B}\partial_r),
\end{equation}
\begin{equation}
X_2=\frac{F}{\psi}(A'\partial_t-\dot{A}\partial_r).
\end{equation}
The Lie bracket of these vector fields is
\begin{equation}
[X_1,X_2]=\frac{Ft(\dot{A}E'-A'\dot{E})}{E\psi}X_1
+\frac{E(\dot{B}F'-B'\dot{F})}{F\psi}X_2.
\end{equation}
For its closedness, we must have
$\frac{F(\dot{A}E'-A'\dot{E})}{E\psi}=a_1$ and
$\frac{E(\dot{B}F'-B'\dot{F})}{F\psi}=a_2$, where $a_1$ and $a_2$
are constants. From here we have either (i) $a_1\neq 0,~a_2=0$, or
(ii) $a_1=0,~ a_2\neq 0$ or (iii) $a_1=0=a_2$. The first two
possibilities contradict the assumption that $\psi\neq 0$. This
shows that the third possibility closes the Lie algebra. Thus we
have
\begin{eqnarray}
C'=\frac{B^2}{A^2A'^2+B^2\dot{A}^2}[\{A'(\ddot{A}-\dot{A}^2)
-\dot{A}(\dot{A}'-A'\dot{B})\}C\nonumber\\
+\{A'(\dot{A}'-\dot{A}A')-\dot{A}(A''-A'B')\}D],
\end{eqnarray}
\begin{eqnarray}
\dot{D}=-\frac{A^2}{A^2A'^2+B^2\dot{A}^2}[\{A'(\ddot{A}-\dot{A}^2)
-\dot{A}(\dot{A}'-A'\dot{B})\}C\nonumber\\
+\{A'(\dot{A}'-\dot{A}A')-\dot{A}(A''-A'B')\}D],
\end{eqnarray}
along with the compatibility constraint in the components $T_0$
and $T_1$ of the energy-momentum tensor given by
\begin{equation}
(\ln\frac{A'}{\dot{A}}e^{A-B})'
(\ln\frac{\dot{B}}{B'}e^{B-A}\dot{)}
-(\ln\frac{A'}{\dot{A}}e^{A-B}\dot{)}
(\ln\frac{\dot{B}}{B'}e^{B-A})'=0.
\end{equation}

\section{Matter Collineations in the Degenerate Case}

In this section only those cases will be considered for which the
energy-momentum tensor is degenerate, i.e., $\det(T_{ab})=0$.

\subsection{MCs of the Manifolds Admitting Higher Symmetries}

Here we would discuss the MCs of the manifolds admitting higher
symmetries than $SO(3)$. For higher symmetries, all metrics have
$T_{01}=0$ except the case (3) of the last section. Thus we would
discuss the spacetimes for which $T_{01}=0$ and $\det(T_{ab})=0$,
i.e., when at least one of the $T_a$ or their combination is zero.
It can be shown that for $T_1=0,~T_k\neq 0,~k=0,2$ (case (1) of
section 3), we obtain infinite dimensional MCs. The solution for
$T_0=0,~T_l\neq 0,~l=1,2$ (case (2) of section 3) also gives
infinite dimensional MCs. These have been discussed in detail
elsewhere [1]. Here we are interested in exploring the
possibilities of finite MCs.

When $T'_k\neq 0,~(\frac{T_0}{T_2})'\neq 0$, we obtain four MCs
which are the usual KVs of the spherical symmetry. For $T'_k\neq
0,~(\frac{T_0}{T_2})'=0$, we obtain ten independent MCs. These are
\begin{eqnarray}
\xi^0=\beta[(\dot{g}_1\sin\phi-\dot{g}_2\cos\phi)\sin\theta
+\dot{g}_3\cos\theta]+c_0,~~\xi^1=0,\nonumber\\
\xi^2=-(g_1\sin\phi-g_2\cos\phi)\cos\theta
+g_3\sin\theta+c_1\sin\phi-c_2\cos\phi,\nonumber\\
\xi^3=-(g_1\cos\phi+g_2\sin\phi)\csc\theta
+(c_1\cos\phi+c_2\sin\phi)\cot\theta+c_3,
\end{eqnarray}
where $\beta=\frac{T_2}{T_0}\neq 0$, is an arbitrary constant and
the function $g$ satisfies the following constraint
\begin{equation}
\beta\ddot{g}_i(t)-g_i(t)=0.
\end{equation}
The solution for the non-static case can be obtained trivially
which turn out to be the same with different constraints.

\subsection{MCs of the Manifolds Admitting $SO(3)$ as the
Maximal Isometry Group}

The metrics which admit $SO(3)$ as the maximal symmetry group
yield $\xi^2=0=\xi^3$ and the MC equations reduce to six
independent equations which involve the following equation
\begin{equation}
\dot{T}_{22}\xi^0+T'_{22}\xi^1=0.
\end{equation}
This gives rise to the following four cases:\\
(i) $~~~~~\dot{T}_{22}=0,~T'_{22}\neq 0,~~~$ (ii)
$\dot{T}_{22}\neq 0,~T'_{22}=0$,\\ (iii) $~~~\dot{T}_{22}\neq
0,~T'_{22}\neq 0,~~~$ (iv) $\dot{T}_{22}=0,~T'_{22}=0$.
\par \noindent
\par \noindent
If we solve these cases, we may have interesting physical
consequences. These will be discussed somewhere else.

\section{Conclusion}

In this paper, we have attempted to classify the most general
spherically symmetric spacetimes according to their MCs. We have
found a general solution of the MC equations for the
non-degenerate, diagonal and non-diagonal energy-momentum tensor.
Further, we have classified spacetimes admitting higher symmetries
than $SO(3)$ and those which admit $SO(3)$ as the maximal isometry
group for both non-degenerate and degenerate cases. It is found
that for the non-degenerate and degenerate cases, we recover the
earlier known results [1] as a special case. We also obtain some
interesting missing results in the earlier work. It is mentioned
here that MCs found here coincide with RCs but the constraints are
entirely different. The summary of the results can be given below
in the
form of tables. \\

\vspace{0.2cm}

{\bf {\small Table 1.}} {\small MCs of Case (1) for the
Non-degenerate Case admitting Higher Symmetries}

\vspace{0.1cm}

\begin{center}
\begin{tabular}{|l|l|l|}
\hline {\bf Cases} & {\bf MCs} & {\bf Constraints}
\\ \hline 1ai & $4$ & $[\frac{T_0}{\sqrt{T_1}}
(\frac{T'_0}{2T_0\sqrt{T_1}})']'\neq 0$
\\ \hline 1aii & $6$ &
$[\frac{T_0}{\sqrt{T_1}}(\frac{T'_0}{2T_0\sqrt{T_1}})']'=0$
\\ \hline 1bi & $4$ &
$\frac{T_2}{\sqrt{T_1}}(\frac{T'_2}{2T_2\sqrt{T_1}})'+1\neq 0 $
\\ \hline 1bii & $7$ &
$\frac{T_2}{\sqrt{T_1}}(\frac{T'_2}{2T_2\sqrt{T_1}})'+1=0,~
(\frac{T'_2}{\sqrt{T_0T_1T_2}})'\neq 0$
\\ \hline 1biii & $4$ &
$\frac{T_2}{\sqrt{T_1}}(\frac{T'_2}{2T_2\sqrt{T_1}})'+1=0,~
(\frac{T'_2}{\sqrt{T_0T_1T_2}})'\neq 0,~(\frac{T'_0}{T_2})'\neq 0$
\\ \hline 1biv & $10$ & $\frac{T_2}{\sqrt{T_1}}
(\frac{T'_2}{2T_2\sqrt{T_1}})'+1=0,~
(\frac{T'_2}{\sqrt{T_0T_1T_2}})'\neq 0,~(\frac{T'_0}{T_2})'=0$
\\ \hline
\end{tabular}
\end{center}
Notice that MCs for the case (2) are the same as for the case (1)
which can be obtained trivially by using the transformations given
in the section (3).

\vspace{0.2cm}

{\bf {\small Table 2}. }{\small MCs of Cases (4),(5),(6) for the
Non-degenerate Case admitting Higher Symmetries}.

\vspace{0.1cm}

\begin{center}
\begin{tabular}{|l|l|l|}
\hline {\bf Cases} & {\bf MCs} & {\bf Constraints}
\\ \hline 4ai & $10$ & $\frac{T_1}{\sqrt{T_0}}
(\frac{\dot{T_2}}{2T_1\sqrt{T_0}}\dot{)}
-k=0,~\dot{T}_1=0$
\\ \hline 4aii & $10$ &
$\frac{T_1}{\sqrt{T_0}}(\frac{\dot{T_2}}{2T_1\sqrt{T_0}}\dot{)}
-k=0,~\dot{T}_1\neq 0$
\\ \hline 4bi & $7$ &
$\frac{T_1}{\sqrt{T_0}}(\frac{\dot{T_2}}{2T_1\sqrt{T_0}}\dot{)}
-k\neq 0,~\dot{T}_1=0$
\\ \hline 4bii & $6$ &
$\frac{T_1}{\sqrt{T_0}}(\frac{\dot{T_2}}{2T_1\sqrt{T_0}}\dot{)}-k\neq
0,~\dot{T}_1\neq 0$
\\ \hline
\end{tabular}
\end{center}
It is noted here that the cases (4), (5) and (6a) describes FRW
metrics and (6b) FRW like metrics and have the same MCS in all
cases as given above.

\vspace{0.2cm}

{\bf {\small Table 3}. }{\small MCs for the Non-degenerate Case
admitting $SO(3)$ as the Maximal Symmetry}.

\vspace{0.1cm}

\begin{center}
\begin{tabular}{|l|l|l|}
\hline {\bf Cases} & {\bf MCs} & {\bf Constraints}
\\ \hline ia & No Proper & $\dot{T}_2=0,~T'_2\neq 0,~\dot{T}_1\neq 0$
\\ \hline ib &Infinite No. of MCs&
$\dot{T}_2=0,~T'_2\neq 0,~\dot{T}_1=0,~\dot{T}_0\neq 0 $
\\ \hline iia & No Proper &
$\dot{T}_2\neq 0,~T'_2=0,~T'_0\neq 0$
\\ \hline iib & Infinite No. of MCs &
$\dot{T}_2\neq 0,~T'_2=0,~T'_0=0,~T'_1\neq 0$
\\ \hline iiia & No Proper &$
\begin{array}{c}
\dot{T}_2\neq 0,~T'_2\neq 0,~
T_0T'_2[\frac{T'_1\dot{T_2}-\dot{T_1}T'_2}{2\sqrt{T_1}\dot{T_2}}
+\{\ln(\frac{\dot{T_2}}{T'_2})\}']\\
+T_1\dot{T_2}[\frac{T'_0\dot{T_2}-\dot{T_0}T'_2}{2\sqrt{T_0}T'_2}
+\{\ln(\frac{\dot{T_2}}{T'_2})\dot{\}}]\neq 0
\end{array}
$ \\ \hline iiib & One Proper & $
\begin{array}{c}
\dot{T}_2\neq 0,~T'_2\neq 0,~
T_0T'_2[\frac{T'_1\dot{T_2}-\dot{T_1}T'_2}{2\sqrt{T_1}\dot{T_2}}
+\{\ln(\frac{\dot{T_2}}{T'_2})\}']\\
+T_1\dot{T_2}[\frac{T'_0\dot{T_2}-\dot{T_0}T'_2}{2\sqrt{T_0}T'_2}
+\{\ln(\frac{\dot{T_2}}{T'_2})\dot{\}}]=0,\\
(\frac{T'_0\dot{T_2}-\dot{T_0}T'_2}{2\sqrt{T_0}\dot{T_2}})'
=[\frac{T'_2}{\dot{T_2}}\{\frac{\dot{T_1}T'_2
-T'_1\dot{T_2}}{2\sqrt{T_1}T'_2}
-(\frac{\dot{T_2}}{T'_2})'\}\dot{]}
\end{array}
$\\ \hline iiic & Infinite No. of MCs & $
\begin{array}{c}
\dot{T}_2\neq 0,~T'_2\neq 0,~
T_0T'_2[\frac{T'_1\dot{T_2}-\dot{T_1}T'_2}{2\sqrt{T_1}\dot{T_2}}
+\{\ln(\frac{\dot{T_2}}{T'_2})\}']\\
+T_1\dot{T_2}[\frac{T'_0\dot{T_2}-\dot{T_0}T'_2}{2\sqrt{T_0}T'_2}
+\{\ln(\frac{\dot{T_2}}{T'_2})\dot{\}}]=0,\\
(\frac{T'_0\dot{T_2}-\dot{T_0}T'_2}{2\sqrt{T_0}\dot{T_2}})'
\neq[\frac{T'_2}{\dot{T_2}}\{\frac{\dot{T_1}T'_2
-T'_1\dot{T_2}}{2\sqrt{T_1}T'_2}
-(\frac{\dot{T_2}}{T'_2})'\}\dot{]}
\end{array}
$\\ \hline iva & One Proper & $
\begin{array}{c}
\dot{T}_2=0,~T'_2=0,~\dot{T}_0T'_1-T'_0\dot{T_1}=0,\\T'_0\neq
0,~[\frac{T_1}{T_0}(\frac{\dot{T}_1}{T'_0}\dot{)}\dot{]}=0
\end{array}
$ \\ \hline ivb & One Proper & $
\begin{array}{c}
\dot{T}_2=0,~T'_2=0,~\dot{T}_0T'_1-T'_0\dot{T_1}=0,~T'_0=0,\\
T'_1\neq
0,~[\frac{T_1}{T_0}(\frac{\dot{T}_1}{T'_0}\dot{)}\dot{]}=0
\end{array}
$ \\ \hline ivc & Infinite No. MCs & $
\begin{array}{c}
\dot{T}_2=0,~T'_2=0,~\dot{T}_0T'_1-T'_0\dot{T_1}=0,~T'_0=0,\\
T'_1=0=\dot{T}_1,~\dot{T}_0\neq 0
\end{array}
$ \\ \hline
\end{tabular}
\end{center}

\vspace{0.2cm}

{\bf {\small Table 4}. }{\small MCs of Degenerate Case admitting
Higher Symmetries}.

\vspace{0.1cm}

\begin{center}
\begin{tabular}{|l|l|l|}
\hline {\bf Cases} & {\bf MCs} & {\bf Constraints}
\\ \hline * & $4$ & $T'_k\neq 0,~(\frac{T_0}{T_1})'\neq 0$
\\ \hline ** & $10$ &
$T'_k\neq 0,~(\frac{T_0}{T_1})'=0$
\\ \hline
\end{tabular}
\end{center}
It can be seen from the above tables that each case has different
constraints on the energy-momentum tensor. It would be interesting
to solve these constraints or at least examples should be
constructed to check the dimensions of the MCs. We are able to
classify MCs of the spacetimes with $SO(3)$ as the maximal
isometry group only for the non-degenerate case. However, it needs
to be completed for the degenerate case. Also, the case (3) of the
section (3) admitting higher symmetries is kept open. These would
be discussed in a separate work.

\newpage
\renewcommand{\theequation}{A\arabic{equation}}
\setcounter{equation}{0}
\section*{Appendix A}

The surviving components of the Ricci tensor are
\begin{eqnarray}
R_{00}&=&\frac{1}{4}e^{\nu-\mu}(2\nu''+\nu'^2-\nu'\mu'
+2\nu'\lambda')-\frac{1}{4}(2\ddot{\mu}+\dot{\mu}^2
-\dot{\nu}\dot{\mu}+4\ddot{\lambda}
+2\dot{\lambda}^2-2\dot{\nu}\dot{\lambda}),\nonumber \\
R_{01}&=&-\frac{1}{2}(2\dot{\lambda}'+\dot{\lambda}\lambda'
-\nu'\dot{\lambda}-\dot{\mu}\lambda'), \nonumber \\
R_{11}&=&\frac{1}{4}e^{\mu-\nu}(2\ddot{\mu}+\dot{\mu}^2
-\dot{\nu}\dot{\mu}+2\dot{\mu}\dot{\lambda})
-\frac{1}{4}(2\nu''+\nu'^2-\nu'\mu'+4\lambda''+2\lambda'^2
-2\mu'\lambda'), \nonumber \\
R_{22}&=&\frac{1}{4}e^{\lambda-\nu}(2\ddot{\lambda}
+2\dot{\lambda}^2 -\dot{\nu}\dot{\lambda}+\dot{\mu}\dot{\lambda})
-\frac{1}{4}e^{\lambda-\mu}(2\lambda''+2\lambda'^2-\mu'\lambda'
+\nu'\lambda')+1,\nonumber \\
R_{33}&=&R_{22}\sin^2\theta.
\end{eqnarray}
The Ricci scalar is given by
\begin{eqnarray}
R&=&\frac{1}{2}e^{-\mu}(2\nu''+\nu'^2-\nu'\mu'+2\nu'\lambda'
-2\mu'\lambda'+3\lambda'^2+4\lambda'')-2e^{-\lambda}\nonumber\\
&-&\frac{1}{2}e^{-\nu}(2\ddot{\mu}+\dot{\mu}^2-\dot{\nu}\dot{\mu}
-2\dot{\nu}\dot{\lambda}+2\dot{\mu}\dot{\lambda}
+3\dot{\lambda}^2+4\ddot{\lambda}).
\end{eqnarray}
Using Einstein field equations (1), the non-vanishing components
of energy-momentum tensor $ T_{ab} $ are
\begin{eqnarray}
T_{00}&=&\frac{1}{4}(\dot{\lambda}^2+2\dot{\mu}\dot{\lambda})
-\frac{1}{4}e^{\nu-\mu}(4\lambda''+3\lambda'^2-2\mu'\lambda')
+e^{\nu-\lambda},~~T_{01}=R_{01},\nonumber \\
T_{11}&=&\frac{1}{4}(\lambda'^2+2\nu'\lambda')
-\frac{1}{4}e^{\mu-\nu}(4\ddot{\lambda}+3\dot{\lambda}^2
-2\dot{\nu}\dot{\lambda})-e^{\mu-\lambda},\nonumber \\
T_{22}&=&\frac{1}{4}e^{\lambda-\mu}(2\nu''+\nu'^2-\nu'\mu'
+\nu'\lambda'-\mu'\lambda'+\lambda'^2+2\lambda'')\nonumber\\
&-&\frac{1}{4}e^{\lambda-\nu}(2\ddot{\mu}+\dot{\mu}^2
-\dot{\nu}\dot{\mu}-\dot{\nu}\dot{\lambda}+\dot{\mu}\dot{\lambda}
+\dot{\lambda}^2+2\ddot{\lambda}),\nonumber \\
T_{33}&=&T_{22}\sin^2\theta.
\end{eqnarray}

\renewcommand{\theequation}{B\arabic{equation}}
\setcounter{equation}{0}
\section*{Appendix B}

The non-vanishing components of the Ricci tensor for FRW
spacetimes are given by
\begin{eqnarray}
R_0&=&-3\frac{\ddot{R}}{R}, \nonumber \\
R_1&=&\frac{(R^3\ddot{)}}{3R}-2k, \nonumber \\
R_2&=&R_1\Sigma ^2(k,r),\nonumber \\
R_3&=&R_2\sin^2\theta,
\end{eqnarray}
where
\begin{eqnarray*}
\Sigma(k,r)&=&\sin r,~~~~~for~~~k=1, \\&=&r,~~~~~~~~~for~~~k=0,\\
&=&\sinh r,~~~for~~~k=-1.
\end{eqnarray*}
The Ricci scalar is given by
\begin{equation}
R=-\frac{6}{R^2}(R\ddot{R}+\dot{R}^2-k).
\end{equation}
Now, the surviving components of energy-momentum tensor for FRW
spacetimes are given by
\begin{eqnarray}
T_0&=&\frac{3}{R^2}(\dot{R}^2-k),\nonumber \\
T_1&=&-(2R\ddot{R}+\dot{R}^2)+k,\nonumber \\
T_2&=&T_1\Sigma^2(k,r),\nonumber \\
T_3&=&T_2\sin^2\theta.
\end{eqnarray}

\renewcommand{\theequation}{C\arabic{equation}}
\setcounter{equation}{0}
\section*{Appendix C}

Linearly independent KVs associated with the FRW spacetimes are
given by [24] for $k=1$
\begin{eqnarray}
\xi_{(1)}&=&\sin\phi\partial_\theta +\cot\theta\cos\phi\partial_\phi,\nonumber\\
\xi_{(2)}&=&\cos\phi\partial_\theta-\cot\theta\sin\phi\partial_\phi,\nonumber\\
\xi_{(3)}&=&\partial_\phi\nonumber\\
\xi_{(4)}&=&(\sin\theta\partial_r+\cot
r\cos\theta\partial_\theta)\sin\phi
+\cot r\csc\theta\cos\phi\partial_\phi,\nonumber\\
\xi_{(5)}&=&(\sin\theta\partial_r+\cot
r\cos\theta\partial_\theta)\cos\phi
-\cot r\csc\theta\sin\phi\partial_\phi,\nonumber\\
\xi_{(6)}&=&\cos\theta\partial_r -\cot r\sin\theta\partial_\theta.
\end{eqnarray}
For $k=0$, we have
\begin{eqnarray}
\xi_{(1)}&=&\sin\phi\partial_\theta +\cot\theta\cos\phi\partial_\phi,\nonumber\\
\xi_{(2)}&=&\cos\phi\partial_\theta-\cot\theta\sin\phi\partial_\phi,\nonumber\\
\xi_{(3)}&=&\partial_\phi\nonumber\\
\xi_{(4)}&=&(\sin\theta\partial_r
+\frac{1}{r}\cos\theta\partial_\theta)\sin\phi
+\frac{1}{r}\csc\theta\cos\phi\partial_\phi,\nonumber\\
\xi_{(5)}&=&(\sin\theta\partial_r
+\frac{1}{r}\cos\theta\partial_\theta)\cos\phi
-\frac{1}{r}\csc\theta\sin\phi\partial_\phi\nonumber\\
\xi_{(6)}&=&(\cos\theta\partial_r
-\frac{1}{r}\sin\theta\partial_\theta).
\end{eqnarray}
For $k=-1$
\begin{eqnarray}
\xi_{(1)}&=&\sin\phi\partial_\theta +\cot\theta\cos\phi\partial_\phi,\nonumber\\
\xi_{(2)}&=&\cos\phi\partial_\theta-\cot\theta\sin\phi\partial_\phi,\nonumber\\
\xi_{(3)}&=&\partial_\phi\nonumber\\
\xi_{(4)}&=&(\sin\theta\partial_r+\coth
r\cos\theta\partial_\theta)\sin\phi
+\coth r\csc\theta\cos\phi\partial_\phi,\nonumber\\
\xi_{(5)}&=&(\sin\theta\partial_r+\coth
r\cos\theta\partial_\theta)\cos\phi
-\coth r\csc\theta\sin\phi\partial_\phi,\nonumber\\
\xi_{(6)}&=&\cos\theta\partial_r -\coth
r\sin\theta\partial_\theta.
\end{eqnarray}

\newpage

\begin{description}
\item  {\bf Acknowledgments}
\end{description}

I would like to thank Ministry of Science and Technology (MOST),
Pakistan for providing postdoctoral fellowship at University of
Aberdeen, UK. I also appreciate the comments given by Prof. G.S.
Hall during its write up.

\vspace{2cm}

{\bf \large References}

\begin{description}

\item{[1]} Sharif, M. and Sehar Aziz: Gen Rel. and Grav. {\bf
35}(2003)1091.

\item{[2]} Katzin, G.H., Levine J. and Davis, W.R.: J. Math. Phys.
{\bf 10}(1969)617;\\ J. Maths. Phys. {\bf 11}(1970)1578.

\item{[3]} Katzin, G.H., Levine J. and Davis, W.R.: Tensor (NS)
{\bf 21}(1970)51;\\ Davis, W.R. and Moss, M.K.: Nuovo Cimento {\bf
65B}(1970)19.

\item{[4]} Katzin, G.H. and Levine, J.: Tensor (NS) {\bf
22}(1971)64;\\ Colloq. Math. {\bf 26}(1972)21.

\item{[5]} Davis, W.R., Green, L.H. and Norris, L.K.: Nuovo
Cimento {\bf 34B}(1976)256;\\ Davis, W.R.: Il Nuovo Cimento {\bf
18}(1977)319.

\item{[6]} Green, L.H., Norris, L.K., Oliver, D.R. and Davis,
W.R.: Gen. Rel. Grav. {\bf 8}(1977)731.

\item{[7]} Stephani, H.: {\it General Relativity: An Introduction
to the Theory of Gravitational Fields} (Cambridge University
Press, 1990).

\item{[8]} Davis, W.R. and Katzin, G.H.: Am. J. Math. Phys. {\bf
30}(1962)750.

\item{[9]} Petrov, A.Z.: {\it Einstein Spaces} (Pergamon, Oxford
University Press, 1969).

\item{[10]} Misner, C.W., Thorne, K.S. and Wheeler, J.A.: {\it
Gravitation} (W.H. Freeman, San Francisco, 1973).

\item{[11]} Kramer, D., Stephani, H., MacCallum, M.A.H. and
Hearlt, E.: {\it Exact Solutions of Einstein's Field Equations}
(Cambridge University Press, 2003).

\item{[12]} Ahsan, Z. and Kang, Tam: J. Maths. {\bf 9}(1978)237.

\item{[13]} Ahsan, Z. and Husain, S.I.: Ann.di. Mat. Pure appl.
{\bf CXXVI}(1980)379.

\item{[14]} Coley, A.A. and Tupper, O.J.: J. Math. Phys. {\bf
30}(1989)2616.

\item{[15]} Hall, G.S., Roy, I. and Vaz, L.R.: Gen. Rel and Grav.
{\bf 28}(1996)299.

\item{[16]} Camc{\i}, U. and Barnes, A.: Class. Quant. Grav. {\bf
19}(2002)393.

\item{[17]} Carot, J. and da Costa, J.: {\it Procs. of the 6th
Canadian Conf. on General Relativity and Relativistic
Astrophysics}, Fields Inst. Commun. 15, Amer. Math. Soc. WC
Providence, RI(1997)179.

\item{[18]} Carot, J., da Costa, J. and Vaz, E.G.L.R.: J. Math.
Phys. {\bf 35}(1994)4832.

\item{[19]} Tsamparlis, M., and Apostolopoulos, P.S.: J. Math.
Phys. {\bf 41}(2000)7543.

\item{[20]} Sharif, M.: Nuovo Cimento {\bf B116}(2001)673;\\
Astrophys. Space Sci. {\bf 278}(2001)447.

\item{[21]} Camc{\i}, U. and Sharif, M.: Gen Rel. and Grav. {\bf
35}(2003)97.

\item{[22]} Camc{\i}, U. and Sharif, M.: Class. Quant. Grav.
(2003).

\item{[23]} Hall, G.S.: Gen. Rel and Grav. {\bf 30}(1998)1099.

\item{[24]} Maartens, R. and Maharaj, S.D.: Class. Quant. Grav.
{\bf 3}(1986)1005.

\end{description}

\end{document}